\documentclass[12pt,twoside]{article}

\usepackage{color}
\definecolor{Red}{cmyk}{0,1,1,0}

\definecolor{Blue}{cmyk}{1,1,0,0}

\usepackage{epsf}
\usepackage{a4wide}
\usepackage{fancyheadings,graphics}
\advance \headheight by 3.0truept       % for 12pt mandatory...

\newlength{\nseparation}
        {\vspace{\nseparation}\par \hrule \end{figure}}

\usepackage{cite}   % collapse adjacent citations

\newcommand{\Bbar}{\,\overline{\!B}}
\newcommand{\bbq}{\ensuremath{B_q\!-\!\Bbar{}_q\,}}

\newcommand{\bra}[1]{\ensuremath{\langle #1 |}}
\newcommand{\ket}[1]{\ensuremath{| #1 \rangle }}

\newcommand{\gqbtf}{\ensuremath{\Gamma (\Bbar{}_q(t) \rightarrow f )}}
\newcommand{\gqtfb}{\ensuremath{\Gamma (B_q(t) \rightarrow \ov{f} )}}

\newcommand{\dm}{\ensuremath{\Delta M}}
\newcommand{\dg}{\ensuremath{\Delta \Gamma}}
\newcommand{\lqcd}{\Lambda_{\rm QCD}}

\newcommand{\ds}{\displaystyle}
\newcommand{\lt}{\left}
\newcommand{\rt}{\right}
\newcommand{\no}{\nonumber}
\newcommand{\nn}{\nonumber \\}
\newcommand{\li}{\mathrm{Li}_2}
\newcommand{\su}{_}

\newcommand{\ov}[1]{\overline{#1}}
\newcommand{\eq}[1]{Eq.~(\ref{#1})}
\newcommand{\imag}{\mathrm{Im}\,}

\newcommand{\gev}{\,\mbox{GeV}}

\newcommand{\epm}[2]{
 \raisebox{-0.5ex}{\shortstack[l]{$\scriptstyle+#1$\\$\scriptstyle-#2$}}}

\begin{document}
\thispagestyle{empty}
\boldmath
\vspace*{-2cm}
\noindent
hep-ph/0307344\hfill PITHA 03/05\\
\mbox{~}\hfill LMU 16/03\\
\mbox{~}\hfill FERMILAB-Pub-03/214-T\\
\mbox{~}\hfill July 2003

\vspace*{1cm}

\centerline{\Large\bf 
CP asymmetry in flavour-specific B decays}
\vspace*{0.3cm}
\centerline{\Large\bf beyond leading logarithms}
\unboldmath

\vspace*{1cm}
\centerline{\sc
   Martin Beneke$^1$,\, Gerhard Buchalla$^2$,} 
\centerline{\sc   Alexander~Lenz$^3$\, and\,  Ulrich Nierste$^4$}

\vspace*{0.5cm}
\centerline{
\parbox{0.89\textwidth}{
\sl $^1$ Institut f\"ur Theoretische Physik E, RWTH Aachen,
         Sommerfeldstra\ss e 28,\\
         \phantom{$^1$} D-52074 Aachen, Germany.\\[2mm]
    $^2$ Ludwig-Maximilians-Universit\"at M\"unchen, Sektion Physik,
         Theresienstra\ss e~37, \\ 
         \phantom{$^1$} D-80333 M\"unchen, Germany.\\[2mm]
    $^3$ Fakult\"at f\"ur Physik, Universit\"at Regensburg, 
        D-93040 Regensburg, Germany.\\[2mm]
    $^4$ Fermi National Accelerator Laboratory, Batavia, 
        IL 60510-500, USA. }}

\vspace*{1cm}
\centerline{\bf Abstract}
\vspace*{0.3cm}
\noindent 
We compute next-to-leading order QCD corrections to the CP asymmetry 
$a_{\rm fs}={\rm Im}(\Gamma_{12}/M_{12})$ in
flavour-specific $B_{d,s}$ decays such as $B_d \to X \ell
\ov{\nu}_\ell$ or $B_s \to D_s^- \pi^+$.  The
corrections reduce the uncertainties associated with the choice of the
renormalization scheme for the quark masses significantly. 
In the Standard Model we predict 
$a^d_{\rm fs}=-(5.0 \pm 1.1)\times 10^{-4}$.
As a by-product we also obtain the width difference in the $B_d$ system at
next-to-leading order in QCD.\\[2mm]

\\[2mm]

%\vfill
%
%\newpage
%\pagenumbering{arabic}
%\setcounter{page}{2}

\section{Preliminaries}
$B_d$ and $B_s$ mesons mix with their antiparticles. The time
evolution of the \bbq\ system (with $q=d$ or $s$) is characterized by
two hermitian $2\times 2$ matrices, the mass matrix $M^q$ and the decay
matrix $\Gamma^q$. The oscillations between the flavour eigenstates
$B_q$ and $\Bbar_q$ involve the three physical quantities
$|M_{12}^q|$, $|\Gamma_{12}^q|$ and $\phi_q=
\arg(-M_{12}^q/\Gamma_{12}^q)$ (see e.g.\ \cite{run2}).  They
are related to the mass and width differences of the $B_q$
system as
\begin{eqnarray}
\dm_q &=& 2\, |M_{12}^q|, 
\qquad \dg_q \; =\; \Gamma^q_L-\Gamma^q_H \; =\; 
        2\, |\Gamma_{12}^q| \cos \phi_q ,  
\end{eqnarray} 
where $\Gamma^q_L$ and $\Gamma^q_H$ denote the widths of the lighter
and heavier mass eigenstate, respectively.  Here and in the following
we neglect tiny corrections of order $|\Gamma_{12}^q/M_{12}^q|^2$.

The CP-violating phase $\phi_q$ can be measured through the CP
asymmetry $a_{\rm fs}^q$ in \emph{flavour-specific} $B_q\to f$ decays,
which means that the decays $\Bbar_q \to f$ and $B_q \to \ov{f}$ are
forbidden \cite{hw}:
\begin{eqnarray}
a^q_{\rm fs} 
     &=& \frac{\gqbtf - \gqtfb}{\gqbtf + \gqtfb} 
     \; = \; \imag \frac{\Gamma_{12}^q}{M_{12}^q} 
    \; = \; \frac{\dg_q}{\dm_q} \tan \phi_q
 . \label{defafs}
\end{eqnarray}
Here $B_q(t)$ and $\Bbar{}_q(t)$ denote mesons which are tagged as a
$B_q$ and $\Bbar_q$ at time $t=0$, respectively. An additional
requirement in \eq{defafs} is the absence of direct CP violation in
$B_q\to f$, which is equivalent to $|\bra{f} B_q \rangle|=
|\bra{\ov{f}} \Bbar_q \rangle| $. For example, $a_{\rm fs}^s$ can be
obtained through $B_s \to D_s^- \pi^+$.  The standard way to access
$a_{\rm fs}^q$ uses $B_q \to X \ell^- \ov{\nu_\ell}$ decays, which
justifies the name \emph{semileptonic CP asymmetry} for $a_{\rm
fs}^q$. The measurement of $a^q_{\rm fs}$ does not require tagging
(see e.g.\ \cite{dfn}). 
A further method to access $a^q_{\rm fs}$ uses the fully inclusive,
tagged $B$ decay asymmetry discussed in \cite{BBD2}. 

$a_{\rm fs}^q$ is small because of two
suppression factors: First $|\Gamma_{12}/M_{12}|=O(m_b^2/M_W^2)$
suppresses $a_{\rm fs}^q$ to the percent level. Second there is a GIM
suppression factor $m_c^2/m_b^2$ reducing $a_{\rm fs}^q$ by another
order of magnitude. This GIM suppression is lifted if new
physics contributes to $\arg M_{12}$. Therefore $a_{\rm fs}^q$ is very
sensitive to new CP phases \cite{run2,llnp}. Up to now, the Standard
Model (SM) prediction for $a_{\rm fs}^q$ was only known in the
leading-logarithmic approximation. The unknown next-to-leading order
(NLO) QCD corrections were identified as the largest theoretical
uncertainty in $a^q_{\rm fs}$ \cite{llnp}. While NLO corrections were
calculated long ago for $M_{12}^q$ \cite{bjw}, only certain portions
of the QCD corrections to $\Gamma_{12}^q$ (relevant to $\dg_s$) were
known so far \cite{bbgln1}. In Sect.~\ref{sec:calc} we compute the
missing pieces of the latter.  Predictions for $a_{\rm fs}^q$ and
$\dg_d$ can be found in Sect.~\ref{sec:phen}. 
%Finally we conclude.

\boldmath
\section{$\Gamma_{12}^q$ at next-to-leading order in QCD}\label{sec:calc}
\unboldmath 

\begin{figure}[t]
\centerline{\epsfxsize=0.6\textwidth \epsffile{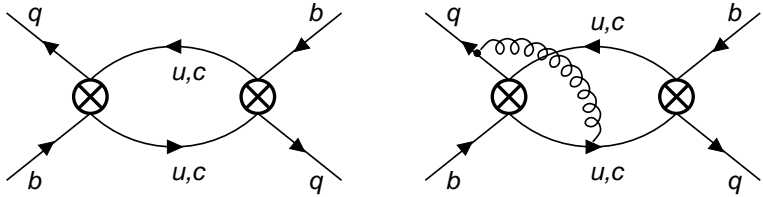}}
\caption{Leading order contribution to $\Gamma_{12}$ (left) and a
sample NLO diagram (right). The crosses denote effective $\Delta B=1$
operators triggering the $b$ decay. The full set of NLO diagrams can
be found in \cite{bbgln1}.}\label{fig:lo}
\end{figure}

In this section we specify the discussion to the case $q=d$ and omit
the index $q$. The ge\-ne\-ra\-li\-za\-tion of our results to 
$\Gamma_{12}^s$ is
straightforward.  $\Gamma_{12}$ is an inclusive quantity stemming from
decays into final states common to $B$ and $\Bbar$.  It can be
computed with the help of the heavy quark expansion (HQE) \cite{hqe}
from diagrams like those in Figure~\ref{fig:lo}. The HQE is a
simultaneous expansion in $\lqcd/m_b$ and $\alpha_s(m_b)$. Corrections
of order $\lqcd/m_b$ to $\Gamma_{12}$ have been calculated in
\cite{BBD1,DHKY} and applied to $a_{\rm fs}$ in \cite{llnp}.

We decompose $\Gamma_{12}$ as
\begin{eqnarray}
\Gamma_{12} &=& - \lt[\, \lambda_c^2 \, \Gamma_{12}^{cc} \; + \; 
        2 \, \lambda_c\,  \lambda_u \, \Gamma_{12}^{uc}\; + \; 
        \lambda_u^{2} \, \Gamma_{12}^{uu} \,  \rt] \label{ga12}
\end{eqnarray}
with the CKM factors $\lambda_i=V_{id}^* V_{ib}$ for $i=u,c,t$.  The
coefficients $\Gamma_{12}^{ab}$, $a,b=u,c$ in \eq{ga12}, which are 
computed from diagrams like those in Figure~\ref{fig:lo}, are positive. 
We present the new NLO expressions for the coefficient
$\Gamma_{12}^{uc}$ in the appendix. $\Gamma_{12}^{cc}$ has
already been given at NLO in \cite{bbgln1}, and  $\Gamma_{12}^{uu}$ 
can be inferred by taking the limit $z\to 0$ in $\Gamma_{12}^{cc}$. 
It is convenient to write
\begin{eqnarray}
\frac{\Gamma_{12}}{M_{12}} 
        &=&  \frac{\lambda_t^2}{M_{12}} \lt[ - \Gamma_{12}^{cc} \; +\;
        \,2\, 
                \lt( \Gamma_{12}^{uc} \, - \, \Gamma_{12}^{cc} \rt)
                \,  \frac{\lambda_u}{\lambda_t} \; 
        +\;  
           \lt( 2\, \Gamma_{12}^{uc}  \, - \, \Gamma_{12}^{cc}   
                \, - \, \Gamma_{12}^{uu}  \rt)  
        \,  \frac{\lambda_u^{2}}{\lambda_t^{2}} \rt]\no\\[1mm]
        &&  \hspace*{-1.5cm} =\, 
            10^{-4} \lt[ \,c_1 + c_2 \frac{B_S^\prime}{B} + c_m \;+\; 
            \lt( a_1 + a_2 \frac{B_S^\prime}{B} + a_m \rt) 
                \,  \frac{\lambda_u}{\lambda_t}  \; + \;  
            \lt( b_1 + b_2 \frac{B_S^\prime}{B} + b_m\rt) 
                \,  \frac{\lambda_u^{2}}{\lambda_t^{2}}\, \rt]. 
  \label{a14}
\end{eqnarray}
Here $B=B(\mu_2=m_b)$ and $B_S^\prime=B_S^\prime(\mu_2=m_b)$ 
parameterize the hadronic matrix elements of the local $\Delta B\!=\!2$
operators $Q$ and $Q_S$:
\begin{eqnarray}
 Q & = &  \ov{q} \gamma_\mu (1-\gamma_5)  b \, 
           \ov{q} \gamma^\mu (1-\gamma_5)  b ,
\qquad 
 Q_S \; = \;  \ov{q} (1+\gamma_5)  b \, \ov{q} (1+\gamma_5)  b , 
 \no \\[1mm]
 \bra{B_d} Q (\mu_2) \ket{\ov{B}{}_d} &=& 
        \frac{8}{3} f_{B_d}^2 M_{B_d}^2 B (\mu_2), \nn
\bra{B_d} Q_S (\mu_2) \ket{\ov{B}{}_d} &=& 
        - \frac{5}{3} f_{B_d}^2 M_{B_d}^2 B_S^{\prime} (\mu_2) 
        \; = \; - \frac{5}{3} f_{B_d}^2 M_{B_d}^2 
  \frac{M_{B_d}^2}{\lt[ \ov{m}_b(\mu_2)+\ov{m}_d(\mu_2) \rt]^2}B_S (\mu_2) .
\label{Bdef}
\end{eqnarray}
The mass $M_{B_d}$ and decay constant $f_{B_d}$ cancel from \eq{a14}.
$B$ and $B_S^\prime$ depend on the scale $\mu_2$ and the
renormalization scheme used in the computation of the matrix elements
in \eq{Bdef}. When combining values for $B_S^\prime/B$ with our results 
for $c_{1,2}$, $a_{1,2}$ and $b_{1,2}$ below, one must verify that they
correspond to the same scheme. Details on the renormalization scheme
used by us can be found in \cite{bbgln1}.  Often the parameter $B_S$
rather than $B_S^\prime$ is chosen to parameterize $\langle
Q_S\rangle$. As shown in \eq{Bdef}, they differ by a factor involving
$\ov{\rm MS}$ masses. $\ov{m}_b(\ov{m}_b)$ is smaller than the pole
mass $m_b$ by roughly 0.4\gev.

For the evaluation of \eq{a14} we also need the SM prediction for
$M_{12}$:
\begin{eqnarray}
M_{12} &=&      \lambda_t^2 \, 
   \frac{G_F^2}{12 \pi^2} M_{B_d} \, \eta_B \, B(\mu_2) b_B (\mu_2) f_{B_d}^2 
    \, M_W^2 S \left( \frac{\ov{m}_t^2}{M_W^2}  \right) 
\label{m12}
\end{eqnarray}
with the QCD factors $\eta_B=0.55$ \cite{bjw} and 
\begin{eqnarray}
b_B(\mu) &=&  \lt[ \alpha_s (\mu ) \rt]^{-6/23} \, 
     \lt[ 1+ \frac{\alpha_s(\mu)}{4 \pi} \frac{5165}{3174}   \rt], 
     \qquad \quad  b_B(m_b) \; = \; 1.52 \pm 0.03 . \no
\end{eqnarray}
Note that results from lattice gauge theory are often quoted for the
scale and scheme invariant parameter $\widehat{B}=b_B(\mu_2) B(\mu_2)$
rather than $B(m_b)$ entering \eq{a14}.

\begin{figure}[t]
\centerline{
\hspace*{-3cm}\epsfxsize=0.8\textwidth \epsffile{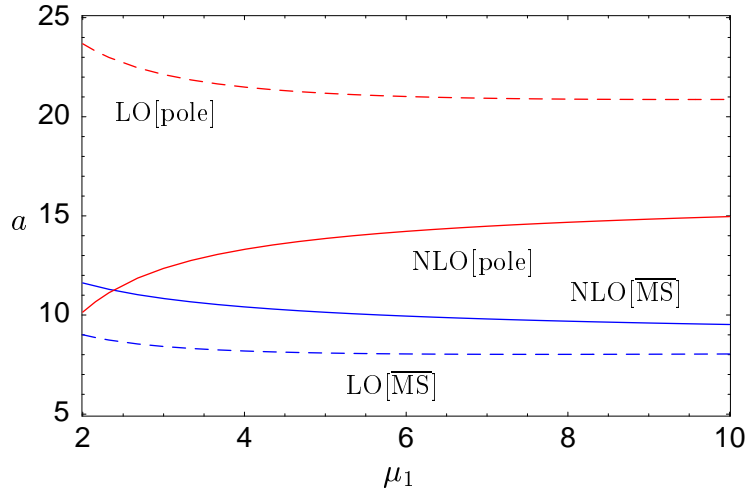} }
\caption{Dependence of $a$ on the scale $\mu_1$. The solid (dashed)
lines show the NLO (LO) results.}\label{fig:sc}
\end{figure}

We use the following input for the physical parameters (where
$\overline{m}_i\equiv \overline{m}_i(\overline{m}_i)$): 
\begin{eqnarray}
\ov{m}_b & = & (4.25 \pm 0.08) \gev, \qquad 
\ov{m}_c \; = \; (1.30 \pm 0.05) \gev, \nn
\alpha_s(M_Z) & = & 0.118 \pm 0.003, \qquad \qquad
\ov{m}_t \; = \; (167 \pm 5) \gev,\nn 
B_S^\prime/B &=& 1.4\pm 0.2, 
\hspace*{2.46cm}m_b^{\rm pow} \; = \; (4.8\pm 0.2) \gev. 
\label{inp}
\end{eqnarray}
The top mass mainly enters the result through $S(\ov{m}_t^2/M_W^2)$ in
\eq{m12}, which evaluates to $S(\ov{m}_t^2/M_W^2)=2.40\pm 0.11$.  In
the power corrections $a_m$, $b_m$, $c_m$ the renormalization scheme
is not fixed, because corrections of order $\alpha_s/m_b$ are
unknown. The expansion parameter of the HQE is the pole mass and we
use $m_b^{\rm pow} = 4.8\pm 0.2\gev$ (and $m_d=7\,{\rm MeV}$)
in $a_m$, $b_m$ and $c_m$. For the determination of
\begin{eqnarray}
a & = & a_1 +  a_2 \frac{B_S^\prime}{B} + a_m  \label{a}
\end{eqnarray}
and the analogously defined quantities $b$ and $c$ we take
$B_S^\prime/B =1.4\pm 0.2$, which covers the range of recent lattice
computations \cite{h}.  We estimate the accuracy of our calculation by
computing the coefficients in two schemes for the quark masses
(pole and $\ov{\rm MS}$), as explained in the appendix.  Further we
vary the renormalization scale $\mu_1$ between one half and twice the 
$b$ quark mass in the corresponding scheme. 
The result is shown in Figure~\ref{fig:sc} for the
coefficient $a$, which is most relevant to $a_{\rm fs}$: While the
dependence on $\mu_1$ is small in both LO and NLO, the scheme
dependence is huge in LO and reduced by roughly a factor of 4 in NLO.
We quote our coefficients for the two schemes and add the errors 
from \eq{inp}, 
and the uncertainty from the $\mu_1$-dependence in quadrature:  
\begin{equation}
\begin{array}[b]{r|r@{\qquad}r@{\qquad}r@{\qquad}r@{\qquad}r}
& \mbox{LO, $\ov{\rm MS}$} & \mbox{LO, pole}  
& \mbox{NLO, $\ov{\rm MS}$} & \mbox{NLO, pole} \\\hline\\[-2mm]
a_1  & 
6.75\epm{0.89}{0.89} &
13.96\epm{1.12}{1.10} &
8.32\epm{1.24}{1.23} &
10.45\epm{0.93}{0.91} \\[2mm]
a_2  &
0.92\epm{0.31}{0.28} &
4.77\epm{1.16}{1.04} &
1.36\epm{0.41}{0.37} &
1.86\epm{1.36}{1.34} \\[2mm]
b_1  &
-0.03\epm{0.01}{0.02} &
-0.31\epm{0.08}{0.10} &
0.00\epm{0.02}{0.02} &
0.10\epm{0.17}{0.17} \\[2mm]
b_2 & 
0.09\epm{0.04}{0.03} &
0.80\epm{0.26}{0.22} &
0.08\epm{0.05}{0.04} &
0.00\epm{0.34}{0.34} \\[2mm]
c_1 & 
-6.60\epm{2.31}{2.32} &
-2.01\epm{3.03}{3.03} &
-3.61\epm{1.32}{1.33} &
-1.01\epm{1.08}{1.08} \\[2mm]
c_2 &
-54.65\epm{7.20}{7.28} &
-61.12\epm{8.08}{8.17} &
-45.54\epm{3.67}{3.77} &
-40.41\epm{6.52}{6.56} \\[2mm]
a_m & 
 0.11\epm{0.06}{0.06} &
0.63\epm{0.31}{0.30} &
0.11\epm{0.06}{0.06} &
0.65\epm{0.32}{0.31} \\[2mm]
% -- &
% -- \\[2mm]
b_m & 
0.03\epm{0.02}{0.02} &
0.23\epm{0.12}{0.11} &
0.03\epm{0.02}{0.02} &
0.24\epm{0.12}{0.12} \\[2mm]
% -- &
%  -- \\[2mm]
c_m &
22.08\epm{9.06}{9.40} &
21.93\epm{8.95}{9.29} &
22.45\epm{9.22}{9.57} &
22.32\epm{9.12}{9.46} \\
% -- &
% -- \\
\end{array} \label{cfnum}
\end{equation}
In the case of $a_m,\ldots, c_m$ the difference between the LO and NLO
columns stems solely from the QCD factor $\eta_B$. 
The reduction of the scheme dependence of $a_1,\ldots, c_2$
is evident from the comparison of the last two
columns with the first two ones.

Our final values
for $a$, $b$, and $c$ are at NLO (LO results in parentheses):
\begin{eqnarray}
a  &=& { 12.0\pm 2.4} \hspace{2.93ex}\qquad\quad(14.7\pm 6.7)\nonumber\\
b  &= & { 0.2\pm 0.1} \hspace{3.98ex}\qquad\quad(0.6\pm 0.5)\nonumber\\
c  &= & -40.1\pm 15.8 \qquad\quad(-63.3\pm 15.6)
\label{abcnum}
\end{eqnarray}
They have been obtained by averaging the results in the
pole scheme and the $\ov{\rm MS}$ scheme for central values
of the input parameters. The error from scheme dependence
was taken to be half the difference between the results
in the two schemes. The errors quoted in \eq{abcnum} were obtained
by combining in quadrature the latter error with the uncertainties
in the $\ov{\rm MS}$ scheme
from scale dependence ($\mu_1$), $\ov{m}_c$, $\ov{m}_b$, $\ov{m}_t$,
$\alpha_s(M_Z)$, $B'_S/B$, and the $b$-mass in the power corrections. 

In order to understand the size of the coefficients $a$, $b$, $c$
at leading and next-to-leading order and the impact of various
uncertainties, it is instructive to expand in the small parameter
$z=m^2_c/m^2_b\sim 0.1$. The leading terms in this expansion
behave as follows: 
\begin{equation}
\begin{array}{r|c|c|c|c|c|c|c}
    & a_1 & a_2 & b_1, b_2 & c_1, c_2 & a_m & b_m & c_m \\
\hline
 & & & & & & & \\[-2mm]
{\rm LO}  & z & z^2 & z^3 & 1 & z^2 &  z^3 & 1 \\[2mm]
\hline
 & & & & & & & \\[-2mm]
{\rm NLO} & \ \alpha_s z, \alpha_s z\ln z\ & \ \alpha_s z \ & 
           \  \alpha_s z^2 \ & \ \alpha_s \  & - & - & -
\end{array}
\label{abczz}
\end{equation}
Here we have displayed the coefficients $a_i$, $b_i$ and $c_i$
separately, indicating the leading order terms and the
NLO corrections. 

In the SM the CP asymmetry $a_{\rm fs}$ does not depend on 
$c_i$, but only on $a_i$ and $b_i$, on which we shall focus 
for the moment. 
Both $a$ and $b$ exhibit an interesting pattern of GIM suppression,
which leads to a pronounced hierarchy among the different contributions.
All of the coefficients of $a_{\rm fs}$ have to vanish as $z\to 0$.
The dominant term is $a_1$, while $a_2$ is suppressed by one,
$b_{1,2}$ even by two additional powers of $z$ at LO. 
{ 
This strong
hierarchy is alleviated at NLO, where the $z^2$ and $z^3$ terms
receive corrections of
order $\alpha_s z$ and $\alpha_s z^2$. Hence they} are still 
parametrically smaller than
$a_1$, which remains the most important coefficient.
As a consequence of this pattern, the coefficients $b_{1,2}$ get
{ larger} relative corrections at NLO, but { remain 
strongly suppressed} in comparison to $a_1$. This suppression is also not
changed by the power corrections $b_m$.
{ Thus $b$ has}
only a minor impact on $a_{\rm fs}$. An additional welcome feature
is the suppression of $a_2$, which considerably reduces the
dependence on the hadronic matrix elements $B'_S/B$. We
emphasize that the dominant term $a_1$ is free of hadronic
uncertainties since the matrix element $B$ in $\Gamma_{12}$
cancels against the identical quantity in $M_{12}$. It can be
seen from \eq{abczz} that power corrections to $a$ are
suppressed by an additional factor of $z$.
As a result of all these properties, $a_{\rm fs}$ is quite accurately 
known in the SM, once the NLO QCD effects are taken into account.
Note that the latter are important to eliminate the sizable
scheme ambiguity of the leading order calculation.
We remark that the $\alpha_s z\ln z$ term in $a_1$
is peculiar to the choice of pole masses 
$z=m^2_{c,pole}/m^2_{b,pole}$, which at one-loop order
is equivalent to $z=\ov{m}^2_c(\ov{m}_c)/\ov{m}^2_b(\ov{m}_b)$.
Expressing the results in terms of 
$\ov{z}=\ov{m}^2_c(\ov{m}_b)/\ov{m}^2_b(\ov{m}_b)$, the $z\ln z$ term is
eliminated. As discussed in \cite{bbgln2} the absence of these
terms holds to all orders in $\alpha_s$. 
Finally, at NLO the overall uncertainty in $a$ and $b$ comes predominantly
from $\ov{m}_c$ and from the residual scheme dependence.

The situation is different for $c$, which is enhanced relative to
$a$, $b$. Here sizable uncertainties are still present at NLO
from the dependence on $B'_S/B$, power corrections and, to a lesser
extent, also from residual scale and scheme dependence.
The parameter $c$ enters the width difference $\Delta\Gamma_d$
and, in general, the expression for $a_{\rm fs}$ in the presence
of new physics. In these cases one has larger theoretical
uncertainties than in the SM analysis of $a_{\rm fs}$.

\section{Phenomenology}\label{sec:phen}

In the SM the CP asymmetry for the $B_d$ system reads
\begin{eqnarray}
a_{\rm fs}^d \; = \; \imag \frac{\Gamma_{12}}{M_{12}} 
        &=&   \lt[ \, a 
                \, \imag \frac{\lambda_u}{\lambda_t} \; 
        + \; 
          b \, \imag \frac{\lambda_u^{2}}{\lambda_t^{2}} \, \rt] \, 10^{-4},
        \label{des}
\end{eqnarray}
where $a$ and $b$ are given in \eq{abcnum}. 
In terms of Wolfenstein parameters $\bar\rho$ and $\bar\eta$
the CKM quantities in \eq{des} are
\begin{equation}\label{lult}
\frac{\lambda_u}{\lambda_t}=
\frac{1-\bar\rho-i \bar\eta}{(1-\bar\rho)^2+\bar\eta^2}-1
=\frac{\cos\beta -i \sin\beta}{R_t}-1
\end{equation}
\begin{equation}\label{imlult}
{\rm Im}\frac{\lambda_u}{\lambda_t}=-\frac{\sin\beta}{R_t}\, ,
\qquad\qquad
{\rm Im}\left(\frac{\lambda_u}{\lambda_t}\right)^2=
\frac{2\sin\beta}{R_t}-\frac{\sin 2\beta}{R^2_t}
\end{equation}
%\begin{equation}\label{imlult}
%{\rm Im}\frac{\lambda_u}{\lambda_t}=
%-\frac{\bar\eta}{(1-\bar\rho)^2+\bar\eta^2}=-\frac{\sin\beta}{R_t}
%\end{equation}
%\begin{equation}\label{imlult2}
%{\rm Im}\left(\frac{\lambda_u}{\lambda_t}\right)^2=
%\frac{2\bar\eta}{(1-\bar\rho)^2+\bar\eta^2}
%   \left(1-\frac{1-\bar\rho}{(1-\bar\rho)^2+\bar\eta^2}\right)=
%\frac{2\sin\beta}{R_t}-\frac{\sin 2\beta}{R^2_t}
%\end{equation}
where $\beta=\arg(-\lambda_t/\lambda_c)$ and
$R_t\equiv\sqrt{(1-\bar\rho)^2+\bar\eta^2}$ are one angle and
one side of the usual unitarity triangle.

\begin{figure}[t]
\centerline{
\hspace*{-3cm}\epsfxsize=0.8\textwidth \epsffile{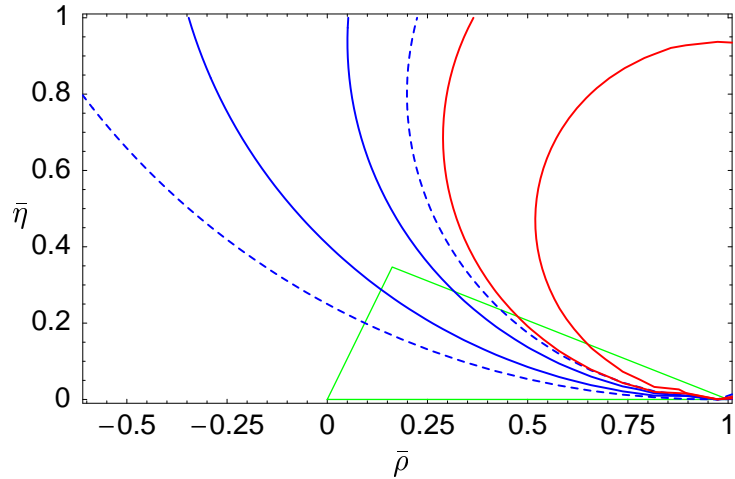} }
\caption{Constraints in the ($\bar\rho,\bar\eta$) plane
implied by given values of the CP asymmetry $a^d_{\rm fs}$.
The area between the solid pair of curves on the right represents
the theoretical uncertainty at NLO, assuming $a^d_{\rm fs}=-10^{-3}$.
Similarly, the curves on the left indicate the uncertainty
for $a^d_{\rm fs}=-5\times 10^{-4}$ both at NLO (solid) and at
LO (dashed). The currently favoured solution for the unitarity
triangle is also shown.}\label{fig:ckm}
\end{figure}

A future measurement of $a^d_{\rm fs}$ will allow us to
constrain $\bar\rho$ and $\bar\eta$ within the SM
using the theoretical values for $a$ and $b$. This is illustrated in
Figure~\ref{fig:ckm}.

Using \eq{abcnum} and \cite{CKMWS} 
\begin{equation}\label{rtbeta}
R_t=0.91\pm 0.05\, , \qquad\qquad  \beta=(22.4\pm 1.4)^\circ
\end{equation}
we predict for $a^d_{\rm fs}$ in the SM
\begin{equation}\label{adfsnum}
 a^d_{\rm fs}= -(5.0 \pm 1.1)\times 10^{-4}
\end{equation}
This result is entirely dominated by the $a$-term in \eq{des}
since the small contribution from $b$ is further suppressed by
its CKM coefficient, which is small for standard CKM parameters.

Our results can also be applied to the case of $B_s$ mesons,
where \eq{des} holds with obvious replacements.
Here the term proportional to $b$ is strongly CKM suppressed
and can be neglected. $SU(3)$ breaking in $a$ is negligible as well
and the result in \eq{abcnum} may be used. We then find ($V_{us}=0.222$)
\begin{equation}\label{asfsnum}
a^s_{\rm fs}=a |V_{us}|^2 R_t \sin\beta \, \times 10^{-4}
 = { (0.21 \pm 0.04)\times 10^{-4} }
\end{equation}

The width difference in the $B_d$ system is given by
$\Delta\Gamma_d/\Delta M_d= -{\rm Re}(\Gamma_{12}/M_{12})$.
The real part of $\Gamma_{12}/M_{12}$ can be found using
Eqs.~(\ref{a14}), (\ref{abcnum}), (\ref{lult}) and (\ref{rtbeta}).
It turns out that for the parameters in \eq{rtbeta} the $c$-term
yields the full result to within about $2\%$. In view of the
large uncertainty of $c$, the contributions from $a$ and $b$
can be safely neglected. We then obtain the SM prediction
\begin{equation}\label{delgamd}
\frac{\Delta\Gamma_d}{\Delta M_d}=(4.0\pm 1.6)\times 10^{-3}\, ,
\qquad\qquad \frac{\Delta\Gamma_d}{\Gamma_d}=(3.0\pm 1.2)\times 10^{-3}
\end{equation}
where the second expression follows with 
the experimental value $\Delta M_d/\Gamma_d=0.755$. This result 
for $\Delta\Gamma_d/\Gamma_d$ is in agreement with \cite{run2,DHKY}.
To the extent that $SU(3)$ breaking in the ratio of bag factors
$B'_S/B$ can be neglected, the number for $\Delta\Gamma/\Delta M$
in \eq{delgamd} applies to the $B_s$ system as well.

The effects of new physics in $M_{12}$ on $a^d_{\rm fs}$ have been
discussed in \cite{llnp}. If magnitude and phase of
$M_{12}$ are parameterized as 
\begin{equation}\label{m12np}
M_{12}=r^2_d e^{2i\theta_d} M^{\rm SM}_{12}
\end{equation}
one obtains \cite{llnp}
\begin{equation}\label{afsnp}
a^d_{\rm fs}=-{\rm Re}\left(\frac{\Gamma_{12}}{M_{12}}\right)_{\rm SM}
\frac{\sin 2\theta_d}{r^2_d}+
{\rm Im}\left(\frac{\Gamma_{12}}{M_{12}}\right)_{\rm SM}
\frac{\cos 2\theta_d}{r^2_d}
\end{equation}
Since the real part of $\Gamma_{12}/M_{12}$ in the SM is much
larger than the imaginary part, $a_{\rm fs}$ is particularly
sensitive to new physics. In this more general context our
results can also be used. However, it has to be kept in mind
that the SM analysis leading to \eq{rtbeta} may no longer
be true in the presence of new physics and the determination
of CKM quantities then needs to be modified.

\vspace*{0.5cm}

To summarize, we have computed the CP violating observables
$a^q_{\rm fs}$ at next-to-leading order in QCD. 
We include the effect of penguin
operators in the weak Hamiltonian and the power corrections
of relative order $\Lambda_{QCD}/m_b$.
Our SM predictions are given in Eqs.~(\ref{adfsnum}) and (\ref{asfsnum}).
We emphasize that within the heavy-quark expansion 
the $a^q_{\rm fs}$ can be reliably computed in the 
SM as functions of CKM parameters. 
A crucial element is the small sensitivity to hadronic parameters,
which enter only as the ratio $B'_S/B$ and only with a suppression
factor of $z=(m_c/m_b)^2$. After including the NLO
corrections, the theoretical error on $a^q_{\rm fs}$ is reduced to
about $20\%$. This is largely due to a reduction of the scheme ambiguity
in the definition of quark masses by a factor of 4 in comparison
with the LO result. 
The remaining uncertainty is larger for $\Delta\Gamma_q$. 
The result at next-to-leading order
in QCD is given in \eq{delgamd}. 
The measurement of $a^q_{\rm fs}$ is possible using
suitable flavour-specific decay modes of neutral $B$ mesons. If it
can be performed with sufficient accuracy, it will provide a significant
test of the Standard Model. The large sensitivity of $a^q_{\rm fs}$
to new physics is reinforced by the improved theoretical analysis
presented here.

\subsubsection*{Note added}
The topic of this paper has also been addressed by Ciuchini et al.\
\cite{cflmt}, who pointed out an error in an earlier preprint version
of this paper. Our analytical results in \eq{coeff} now agree with
those in Eqs.~(43-45) of \cite{cflmt}. We thank the authors of
\cite{cflmt} for clarifying communication.

\subsubsection*{Acknowledgement}
The work of M.B.\ 
is supported in part by the Bundesministerium f\"ur Bildung und Forschung, 
Project 05~HT1PAB/2, and by the DFG Sonder\-forschungsbereich/Transregio~9 
``Computer-gest\"utzte Theoretische Teilchenphysik''. 

\appendix
\section{NLO coefficients}
Here we collect more detailed results for the coefficients in \eq{ga12}.  
The HQE expresses
$\Gamma_{12}^{ab}$ for the $B_d$ system as
\begin{eqnarray}
\Gamma_{12}^{ab} &=& \frac{G_F^2 m_b^2}{24 \pi} f_{B_d}^2 M_{B_d} 
   \lt[ \lt( F^{ab}(z)+P^{ab}(z) \rt) \frac{8}{3} B 
        - \lt( F_S^{ab}(z)+P_S^{ab}(z) \rt) \frac{5}{3} B_S^\prime \rt]
      + \Gamma_{12,1/m_b}^{ab}.
        \label{deffp}
\end{eqnarray}
The short-distance coefficients $F^{ab}(z)$ contain the contributions
from the $\Delta B\!=\!1$ current-current operators $Q_1$ and $Q_2$.
The NLO results for $F^{cc}(z)$ and $F_S^{cc}(z)$ have been derived in
\cite{bbgln1}, where these coefficients are called $F(z)$ and
$F_S(z)$, respectively. Further $F^{uu}=F^{cc}(0)$ and
$F_S^{uu}=F_S^{cc}(0)$. The coefficients $P(z)$ and $P_S(z)$ contain
the contributions from penguin operators. They come with small
coefficients, which simplifies the NLO calculation \cite{bbgln1}. 

Our new calculation concerns $F^{uc}$, { $F_S^{uc}$, $P^{uc}$ and
$P_S^{uc}$.} We decompose { $F^{uc}$ and $F_S^{uc}$} as in
\cite{bbgln1,bbgln2}:
\begin{eqnarray}
F^{uc} (z) &=& C_1^2 F_{11}^{uc} (z) + C_1 C_2 F_{12}^{uc} (z) + 
          C_2^2 F_{22}^{uc} (z),\nn
F_{ij}^{uc} (z) &=& F_{ij}^{uc,(0)} (z) + \frac{\alpha_s
          (\mu_1)}{4\pi} F_{ij}^{uc,(1)} (z,x_{\mu_1},x_{\mu_2}) 
        + {\cal O} (\alpha_s^2)   
   \label{decfuc}
\end{eqnarray}
with $x_{\mu}=\mu/m_b$ and an analogous notation for $F_{S,ij}^{uc}$.
The $\Delta B\!=\! n$ operators, $n=1,2$, are defined at the scale 
$\mu_n={\cal O}(m_b)$. 
The dependence of $F_{ij}^{uc}$ on $\mu_1$ diminishes
order-by-order in $\alpha_s$. 

Throughout this paper we use the same operator definitions
and renormalization schemes as in \cite{bbgln1}, with one important
addition: In $a_{\rm fs}$ the renormalization scheme of the quark
masses is an important issue and we choose two different schemes
for the computation of the $a_i$, $b_i$, $c_i$  in \eq{a14}. 
For both schemes we take the $\ov{\rm MS}$ masses $\ov{m}_c(\ov{m}_c)$
and $\ov{m}_b(\ov{m}_b)$ as the basic input.
In the first scheme (pole scheme) we express the observables
in terms of $m_b=m_{b,pole}=\ov{m}_b (1+4\alpha_s(\ov{m}_b)/3\pi)$,
using the one-loop relation between pole- and $\ov{\rm MS}$-quark mass.
In this scheme we define the variable $z$ as 
$z=(\ov{m}_c(\ov{m}_c)/\ov{m}_b(\ov{m}_b))^2$, which to one-loop order
is equivalent to the ratio of pole masses squared.
In the second scheme ($\ov{\rm MS}$ scheme) we take $m_b=\ov{m}_b(\ov{m}_b)$
and replace $z$ by 
$\ov{z}=(\ov{m}_c(\ov{m}_b)/\ov{m}_b(\ov{m}_b))^2$, where both
running masses are defined at the scale $\ov{m}_b$.
The results below for the functions $F^{uc,(1)}_{ij}(z)$ are valid in the
pole scheme. The corresponding functions 
$\ov{F}{}^{ab,(1)}_{ij}(\ov{z})$ in the
$\ov{\rm MS}$ scheme are obtained via the relation
\begin{equation}
\ov{F}{}^{ab,(1)}_{ij}(\ov{z}) = F^{ab,(1)}_{ij}(\ov{z}) +
		\frac{32}{3} F^{ab,(0)}_{ij}(\ov{z})
  -8 \ov{z} \ln \ov{z}\, 
	\frac{\partial F^{ab,(0)}_{ij}(\ov{z})}{\partial \ov{z}}\, .
\end{equation}
The coefficients read:
\begin{eqnarray}
F^{uc,(0)}_{11}(z) &=& 3(1-z)^2 (1+\frac{z}{2}) \nonumber \\
F^{uc,(0)}_{12}(z) &=& 2(1-z)^2 (1+\frac{z}{2}) \nonumber \\
F^{uc,(0)}_{22}(z) &=& \frac{1}{2}(1-z)^3 \nonumber \\
F^{uc,(0)}_{S,11}(z) &=& 3(1-z)^2 (1+ 2 z) \nonumber \\
F^{uc,(0)}_{S,12}(z) &=& 2(1-z)^2 (1+ 2 z) \nonumber \\
F^{uc,(0)}_{S,22}(z) &=& -(1-z)^2 (1+ 2 z)
\end{eqnarray}

{\small 
\begin{eqnarray}
\lefteqn{
F^{uc ,(1)}_{11} (z,x_{\mu_1},x_{\mu_2})
 = \left[
16\,{\left( 1 - z \right) }^2\,\left( 2 + z \right) 
 \right] \; \ds \left[\li(z) + \frac{\ln (1 - z)\,\ln (z)}{2}\right]
 \; +} &&  \no \\[2mm]
 &&\ds \left[
-4\,{\left( 1 - z \right) }^2\,\left( 5 + 7\,z \right) 
 \right] \; \ds \ln (1 - z) \; + \;
\ds \left[
-2\,z\,\left( 10 + 14\,z - 15\,z^2 \right) 
 \right] \; \ds \ln (z) \; + \no \\[2mm]
 &&\ds \left[
2\,{\left( 1 - z \right) }^2\,\left( 5 + z \right) 
 \right] \; \ds \ln (x\su {\mu\su 2}) \; + \;
\ds
\frac{\left( 1 - z \right) \,\left( 109 - 113\,z - 104\,z^2 \right) }{6}
\no 
\end{eqnarray}

\begin{eqnarray}
\lefteqn{
F^{uc ,(1)}_{12} (z,x_{\mu_1},x_{\mu_2})
 = \left[
\frac{32\,{\left( 1 - z \right) }^2\,\left( 2 + z \right) }{3}
 \right] \; \ds \left[\li(z) + \frac{\ln (1 - z)\,\ln (z)}{2}\right] 
 \; +} && \no \\[2mm]
 &&\ds \left[
\frac{-\left( {\left( 1 - z \right) }^2\,
      \left( 2 + 33\,z + 94\,z^2 \right)  \right) }{6\,z}
 \right] \; \ds \ln (1 - z) \; + \;
\ds \left[
\frac{-\left( z\,\left( 80 + 69\,z - 126\,z^2 \right)  \right) }{6}
 \right] \; \ds \ln (z) \; + \no \\[2mm]
 &&\ds \left[
-2\,{\left( 1 - z \right) }^2\,\left( 17 + 4\,z \right) 
 \right] \; \ds \ln (x\su {\mu\su 1}) \; + \;
\ds \left[
\frac{4\,{\left( 1 - z \right) }^2\,\left( 5 + z \right) }{3}
 \right] \; \ds \ln (x\su {\mu\su 2}) \; + \no \\[2mm]
 &&\ds
\frac{\left( 1 - z \right) \,\left( -502 + 410\,z + 23\,z^2 \right) }{18}
\no 
\end{eqnarray}

\begin{eqnarray}
\lefteqn{
F^{uc ,(1)}_{22} (z,x_{\mu_1},x_{\mu_2})
 = \left[
\frac{2\,\left( 5 - 8\,z \right) \,\left( 1 - z \right) \,
    \left( 1 + 2\,z \right) }{3}
 \right] \; \ds \left[\li(z) + \frac{\ln (1 - z)\,\ln (z)}{2}\right] 
 \; +} &&  \no \\[2mm]
 &&\ds \left[
\frac{{\left( 1 - z \right) }^2\,\left( 7 + 32\,z^2 + 3\,z^3 \right) }{6\,z}
 \right] \; \ds \ln (1 - z) \; + \;
\ds \left[
\frac{-\left( z\,\left( 62 - 39\,z - 30\,z^2 + 3\,z^3 \right)  \right) }{6}
 \right] \; \ds \ln (z) \; + \no \\[2mm]
 &&\ds \left[
-2\,{\left( 1 - z \right) }^2\,\left( 5 + 4\,z \right) 
 \right] \; \ds \ln (x\su {\mu\su 1}) \; + \;
\ds \left[
\frac{2\,{\left( 1 - z \right) }^2\,\left( 4 - z \right) }{3}
 \right] \; \ds \ln (x\su {\mu\su 2}) \; + \no \\[2mm]
 &&\ds \left[
\frac{\left( 1 - z \right) \,\left( -1 + 4\,z \right) }{3}
 \right] \; \ds {\pi }^2 \; + \;
\ds
\frac{\left( 1 - z \right) \,\left( -136 - 295\,z + 443\,z^2 \right) }{18}
\no 
\end{eqnarray}

\begin{eqnarray}
\lefteqn{
F^{uc ,(1)}_{S,11} (z,x_{\mu_1},x_{\mu_2})
 = \left[
32\,{\left( 1 - z \right) }^2\,\left( 1 + 2\,z \right) 
 \right] \; \ds \left[\li(z) + \frac{\ln (1 - z)\,\ln (z)}{2}\right] 
 \; +} &&  \no \\[2mm]
 &&\ds \left[
-8\,{\left( 1 - z \right) }^2\,\left( 4 + 14\,z - 3\,z^2 \right) 
 \right] \; \ds \ln (1 - z) \; + \;
\ds \left[
-8\,z\,\left( -2 + 23\,z - 21\,z^2 + 3\,z^3 \right) 
 \right] \; \ds \ln (z) \; + \no \\[2mm]
 &&\ds \left[
-32\,{\left( 1 - z \right) }^2\,\left( 1 + 2\,z \right) 
 \right] \; \ds \ln (x\su {\mu\su 2}) \; + \;
\ds
\frac{-4\,\left( 1 - z \right) \,\left( 10 - 23\,z + 31\,z^2 \right) }{3}
\no 
\end{eqnarray}

\begin{eqnarray}
\lefteqn{
F^{uc ,(1)}_{S,12} (z,x_{\mu_1},x_{\mu_2})
 = \left[
\frac{64\,{\left( 1 - z \right) }^2\,\left( 1 + 2\,z \right) }{3}
 \right] \; \ds \left[\li(z) + \frac{\ln (1 - z)\,\ln (z)}{2}\right] 
 \; +} && \no \\[2mm]
 &&\ds \left[
\frac{-4\,{\left( 1 - z \right) }^2\,
    \left( 1 + 15\,z + 47\,z^2 - 12\,z^3 \right) }{3\,z}
 \right] \; \ds \ln (1 - z) \; + \no \\[2mm]
&&
\ds \left[
\frac{-4\,z\,\left( -8 + 93\,z - 87\,z^2 + 12\,z^3 \right) }{3}
 \right] \; \ds \ln (z) \; + \no \\[2mm]
 &&\ds \left[
-16\,{\left( 1 - z \right) }^2\,\left( 1 + 2\,z \right) 
 \right] \; \ds \ln (x\su {\mu\su 1}) \; + \;
\ds \left[
\frac{-64\,{\left( 1 - z \right) }^2\,\left( 1 + 2\,z \right) }{3}
 \right] \; \ds \ln (x\su {\mu\su 2}) \; + \no \\[2mm]
 &&\ds
\frac{2\,\left( 1 - z \right) \,\left( -130 - 37\,z + 107\,z^2 \right) }{9}
\no 
\end{eqnarray}

\begin{eqnarray}
\lefteqn{
F^{uc ,(1)}_{S,22} (z,x_{\mu_1},x_{\mu_2})
 = \left[
\frac{16\,\left( 1 - 4\,z \right) \,\left( 1 - z \right) \,
    \left( 1 + 2\,z \right) }{3}
 \right] \; \ds \left[\li(z) + \frac{\ln (1 - z)\,\ln (z)}{2}\right] 
 \; +} && \no \\[2mm]
 &&\ds \left[
\frac{4\,{\left( 1 - z \right) }^2\,\left( 1 + z \right) \,
    \left( -1 + 13\,z + 3\,z^2 \right) }{3\,z}
 \right] \; \ds \ln (1 - z) \; + \;
\ds \left[
\frac{4\,z\,\left( 2 - 3\,z + 18\,z^2 - 3\,z^3 \right) }{3}
 \right] \; \ds \ln (z) \; + \no \\[2mm]
 &&\ds \left[
-16\,{\left( 1 - z \right) }^2\,\left( 1 + 2\,z \right) 
 \right] \; \ds \ln (x\su {\mu\su 1}) \; + \;
\ds \left[
\frac{32\,{\left( 1 - z \right) }^2\,\left( 1 + 2\,z \right) }{3}
 \right] \; \ds \ln (x\su {\mu\su 2}) \; + \no \\[2mm]
 &&\ds \left[
\frac{8\,\left( 1 - z \right) \,\left( 1 + 2\,z \right) }{3}
 \right] \; \ds {\pi }^2 \; + \;
\ds
\frac{28\,\left( 1 - z \right) \,\left( -5 - 8\,z + 19\,z^2 \right) }{9}
\label{coeff} 
\end{eqnarray}

}
In terms of the function $P(z)$ used in \cite{bbgln1} the penguin
coefficients in \eq{deffp} read $P^{cc}(z)=P(z)$, $P^{uu}=P(0)$ and 
\begin{equation}
P^{uc}(z) \; = \; \frac{P(z)+P(0)}{2} \, + \, 
        \Delta P^{uc},  \qquad 
P_S^{uc}(z) \; = \; \frac{P_S(z)+P_S(0)}{2} \, - \, 8
         \Delta P^{uc} 
\end{equation}
with
\begin{equation}
\Delta P^{uc} \; = \; 
\frac{\alpha_s(\mu_1)}{4\pi} \, C_2^2(\mu_1) \,
        \frac{1-(1+ 2 z) \sqrt{1- 4 z}}{18} \, 
                \lt[ \ln z  \, - \, (1+ 2 z) \sqrt{1- 4 z} \ln \sigma 
                  \, - \, 4 z  \rt] 
\label{puc}  
\end{equation}
and $\sigma=(1-\sqrt{1-4 z})/(1+\sqrt{1-4z})$. $\Delta P^{uc}$ is of
order $z^3$ and numerically negligible.

The power corrections $\Gamma^{ab}_{12,1/m_b}$ were first obtained
for $ab=cc,uu$ in \cite{BBD1} and for $ab=uc$ in \cite{DHKY}.
We have re-computed the case $ab=uc$ here, confirming the results
of \cite{DHKY}.\
In the notation of \cite{BBD1} we find
($\langle\ldots\rangle\equiv\langle\bar B|\ldots|B\rangle$)
\begin{eqnarray}
\Gamma^{uc}_{12,1/m_b} &=& \frac{G^2_F m^2_b}{24\pi M_B} 
 (1-z)^2 \Biggl[ (1+2z)K_2\langle R_0\rangle 
-2(1+2z)(K_1\langle R_1\rangle +K_2 \langle \tilde R_1\rangle)  
\nonumber \\
& & 
-2\frac{1+z+z^2}{1-z}(K_1\langle R_2\rangle +K_2 \langle \tilde R_2\rangle)
-\frac{12 z^2}{1-z} (K_1\langle R_3\rangle +K_2 \langle \tilde R_3\rangle)
\Biggr]\, .
\end{eqnarray}

\end{document}